\documentclass[usenatbib,referee]{mn2e}
\usepackage{graphicx}
\usepackage{epsfig}
\usepackage{color}

\title[On the evolution of rotating accreting WDs and SNe Ia]
{On the evolution of rotating accreting white dwarfs and type Ia supernovae}
\author[B. Wang et al.]
{B. Wang,$^{\rm 1,2}$\thanks{E-mail:wangbo@ynao.ac.cn} S. Justham,$^{\rm 3}$ Z.-W. Liu,$^{\rm 4}$ J.-J. Zhang,$^{\rm 1,2}$  D.-D. Liu$^{\rm 1,2}$ and Z. Han$^{\rm 1,2}$  \\
$^1$Yunnan Observatories, Chinese Academy of Sciences, Kunming 650011, China\\
$^2$Key Laboratory for the Structure and Evolution of Celestial Objects, Chinese Academy of Sciences, Kunming 650011, China\\
$^3$National Astronomical Observatories, Chinese Academy of Sciences, Beijing 100012, China\\
$^4$Argelander-Institut f\"{u}r Astronomie, Auf dem H\"{u}gel 71, D-53121, Bonn, Germany}

\begin{document}
%\date{Accepted. Received}
\date{Accepted 2014 September 9}
\pagerange{\pageref{firstpage}--\pageref{lastpage}} \pubyear{2014}
\maketitle

\label{firstpage}

\begin{abstract}
The potential importance of the angular momentum which is
gained by accreting white dwarfs (WDs) has been increasingly recognized
in the context of type Ia supernova (SN Ia) single-degenerate  model.
The expectation that the spin of the WD can delay the explosion should
help the single-degenerate model to be consistent with the observed
properties of most SNe Ia, in particular by avoiding hydrogen contamination.
In this article, we attempt to study the most prominent single-degenerate
supersoft (WD + MS) channel when the rotation of accreting WDs is considered.
We present a detailed binary population synthesis study to examine the
predicted population of SNe Ia for this channel.  For our standard model,
we find that 77\% of these SNe Ia explode with
WD masses which are low enough to be supported by solid-body rotation
($\leq$$1.5\,\rm M_{\odot}$); this is a substantially higher proportion
than found by previous work. Only 2\% have WD explosion masses
$\geq$2.0$\,\rm M_{\odot}$; these require the initial WD mass to be larger
than 1.0$\,\rm M_{\odot}$. We further discuss the possible origin of the
diversity of SNe Ia from the pre- and post- accretion properties of the
WDs in this population. We also suggest that some SN Ia progenitors with
substantial circumstellar hydrogen, including some apparent type IIn SNe,
might be related to WDs which required support from differential
rotation to avoid explosion, since these can still be accreting from
hydrogen-rich donors with a relatively high mass-transfer rate at the
time of the SN explosion.
\end{abstract}


\section{Introduction} \label{1. Introduction}
Type Ia supernova (SN~Ia) explosions are amongst the most energetic events
observed in the Universe, and they are valuable probes for the study of
cosmic evolution. Empirical correlations which allow the inference of absolute
luminosities from observed lightcurves have enabled SNe~Ia to be used as
distance indicators. These methods, combined with their visibility to cosmological
scales, enabled the determination of the accelerating expansion of the Universe
(e.g. Riess et al.\ 1998; Perlmutter et al.\ 1999).
Even though SNe~Ia have been successfully used as standardizable candles for measuring
cosmological distances, there exists diversity amongst SNe~Ia that is presently
not well understood. Nor do we know how this diversity is linked to the properties of
their progenitors and the explosion mechanism. Understanding the variety which is
contained within the population of SNe Ia should help to constrain their origin
and possible explosion mechanisms. In principle, there might be random reasons
why SNe Ia appear different from each other (e.g. the number of ignition points,
the location of the deflagration-to-detonation transition or symmetry-breaking produced by
aspherical explosions; see, e.g.\ Hillebrandt \& Niemeyer 2000;  R\"{o}pke \& Hillebrandt 2004;
Kasen, R\"{o}pke \& Woosley 2009; Maeda et al.\ 2010; Chen, Han \& Meng 2014). However, the existence of clear
systematic population differences suggests that a significant amount of the
diversity must arise from systematic causes (for recent observational evidence,
see, e.g. Wang \& Wheeler 2008; Howell et al.\ 2009; Sullivan et al.\ 2010; Childress et al.\ 2013;
Wang et al.\ 2012, 2013; Pan et al.\ 2014). These systematic differences seem
likely to be able to be traced to the properties of the progenitors,
either because of a changing mix of qualitatively different
progenitor types or from continuous variation within one progenitor
type or both.

It has been widely accepted that SN Ia explosion occurs when a carbon--oxygen white
dwarf (CO WD) is destroyed in a thermonuclear explosion. Over the past few decades,
two families of models have been proposed to produce CO WDs which reach the conditions
necessary for an explosion, i.e. the single-degenerate (SD) and double-degenerate (DD)
models. Numerous variants of both the SD and DD scenarios exist (for SD models see, e.g.
Whelan \& Iben 1973; Hachisu et al.\ 1996; Li \& van den Heuvel 1997; Yungelson \& Livio 1998; Langer et al.\ 2000;
Han \& Podsiadlowski 2004; Meng, Chen \& Han 2009; Wang et al.\ 2009a; L\"{u} et al. (2009);
Ablimit, Xu \& Li 2014; Claeys et al. 2014; whilst for DD models see, e.g. Webbink 1984; Iben \& Tutukov 1984;
Nelemans et al. 2001; van Kerkwijk, Chang \& Justham 2010; Chen et al. 2012; Toonen, Nelemans \& Portegies Zwart 2012).
Different pieces of evidence can be used to support or cause problems for the SD and DD classes, but
no definitive arguments have excluded either option. For recent reviews on this subject see
Podsiadlowski et al.\ (2008), Howell (2011), Wang \& Han (2012), Parrent, Friesen \& Parthasarathy
(2014) and Maoz, Mannucci \& Nelemans (2014).

In the classic SD model, the WD gains matter from a non-degenerate companion star until
it reaches a critical mass limit -- near to but not coincident with the Chandrasekhar mass,
$M_{\rm Ch}$ -- at which the core properties of the WD are expected to result in runaway explosive
carbon burning (for a calculation of the critical mass see Nomoto et al.\ 1997).\footnote{For the SD model,
however, a vigorous debate is still going on about the process of the mass accretion (see Cassisi,
Iben \& Tornamb\`{e} 1998; Idan, Shaviv \& Shaviv 2012; Newsham, Starrfield \& Timmes 2013).} If the
mass-donating companion is a main sequence (MS) star or a slightly evolved star, the formation
channel is generally referred to as the ``WD + MS channel'' even if the mass-donating star is a
subgiant star; an alternative name for this is the ``supersoft channel'' (e.g.  Li \& van den Heuvel 1997;
Langer et al.\ 2000; Han \& Podsiadlowski 2004). This supersoft channel is the most
commonly-considered variant of the SD model, and the one which we will concentrate on in this
binary population synthesis (BPS) calculations. Additionally, the importance of the rotation of
the accreting WD has been increasingly recognized in the context of the SD model
(e.g. Yoon \& Langer 2004; Justham 2011; Di Stefano, Voss \& Claeys 2011; Hachisu et al. 2012a,b).
Hachisu et al.\ (2012a) provided a clear summary of how considering the spin of the WD may be vital
to help the SD model be consistent with observations of SNe Ia.

In recent years, a few over-luminous SNe Ia have been observed with inferred WD explosion masses
of order $2\,\rm M_{\odot}$. SN 2003fg is the first observed over-luminous event. It was discovered
on 24 April 2003, and observed to be 2.2 times more luminous than a normal one, and the amount of $^{56}$Ni was inferred
to be $1.29\pm0.07\,\rm M_{\odot}$, which would suggest a super-Chandrasekhar mass WD explosion
with $M_{\rm WD}\sim2.1\,\rm M_{\odot}$ (Howell et al. 2006). Following the discovery of SN 2003fg,
three over-luminous events were discovered, i.e. SN~2006gz ($M_{\rm Ni}\sim1.2\,\rm M_{\odot}$;
Hicken et al. 2007), SN~2007if ($M_{\rm WD}\sim2.4\pm0.2\,\rm M_{\odot}$ with $M_{\rm Ni}\sim1.6\pm0.1\,\rm M_{\odot}$;
Scalzo et al.\ 2010) and SN~2009dc ($M_{\rm WD}>2.0\,\rm M_{\odot}$ with $M_{\rm Ni}=1.4-1.7\,\rm M_{\odot}$;
Silverman et al.\ 2011).  One possibility is that super-Chandrasekhar mass WDs produce these
over-luminous SNe~Ia following the mergers of double WD systems (e.g. Howell et al.\ 2006).
If the SD model is to explain the existence of these events, then in these cases the WDs must have been
temporarily prevented from exploding by the effect of differential rotation. For the study of the production
of these events by SD systems see, e.g. Chen \& Li (2009) and Hachisu et al. (2012a,b). Note that models for
the luminosity of these events which do not require super-Chandrasekhar mass WDs have been suggested by
Hillebrandt, Sim \& R\"{o}pke (2007) and Hachinger et al.\ (2012).

Models for SD SN Ia progenitors which include the rotation of WDs find that the SNe should
occur from WDs with a range of final mass. The main theoretical uncertainty is whether
the accreting WDs are typically able to sustain differential rotation. If differential
rotation is easily maintained, then in principle the WD mass distribution at explosion
could extend far above $M_{\rm Ch}$; for extreme differential rotation, Ostriker \& Bodenheimer
(1968) calculated that a WD could be stable to be $\sim$4$\,\rm M_{\odot}$.  However, for more
realistic conditions, Yoon \& Langer (2004) found that some WDs could reach $\sim$2$\,\rm M_{\odot}$;
their adopted angular momentum transport theory required mass-transfer rates in excess of
$\sim$$3.0\times10^{-7}\,\rm M_{\odot}\,\rm yr^{-1}$ to maintain differential rotation.
Hachisu (1986) also suggested the existence of stable equilibrium configurations with
$M_{\rm WD}\geq2\,\rm M_{\odot}$. In contrast, Saio \& Nomoto (2004) and Piro (2008) argued
that angular-momentum transport is likely to be so efficient that the accreting WD will
be forced to approach solid-body rotation. For pure solid-body rotation, the WD stability
limit is $\sim$$1.47\,\rm M_{\odot}$ (Anand 1965; Uenishi, Nomoto \& Hachisu 2003; Saio \& Nomoto 2004).
In this case, the possible mass range of WDs produced by the SD SN Ia models would be
restricted to within $\sim$$0.1\,\rm M_{\odot}$ of the standard ignition mass.

The purpose of this article is to investigate the WD + MS channel systematically when
the effect of rotation on the accreting WDs is considered.
Firstly, we followed the evolution of the WD + MS binary until the WD increases its mass to the maximum,
and thereby obtained the initial and final parameter spaces for the production of
SNe Ia. We then used these results to perform a detailed BPS
approach in order to obtain SN Ia birthrates and delay times, and the final fate of these systems.
This paper is organized as follows. In Section 2, we describe the numerical code for our binary evolution
calculations. The binary evolutionary results are shown in Section 3. We describe the BPS method in
Section 4 and present the BPS results in Section 5. Finally, a discussion is given in Section 6,
and a summary in Section 7.

\section{Numerical code for binary evolution calculations}
In the WD + MS channel, the Roche-lobe filling star is a MS star or a slightly evolved subgiant star. The
mass-donating star transfers some of its matter to the surface of the
WD, which leads to the mass increase of the WD. In the previous works (e.g. Han \& Podsiadlowski 2004;
Meng, Chen \& Han 2009; Wang, Li \& Han 2010), we stopped the evolution of the WD + MS binary
at the moment when the WD increases its mass to 1.378$\,\rm M_{\odot}$ (i.e. the critical
mass limit of non-rotating WDs for carbon ignition; Nomoto et al.\ 1984).
In the present works, however, we further follow the evolution of the WD + MS binary until
the mass-transfer rate, $|\dot{M}_{\rm 2}|$, decreases to a critical rate,
$3.0\times10^{-7}\,\rm M_{\odot}\,\rm yr^{-1}$ ($=\dot{M}_{\rm r}$), since the WD is expected
to be supported by a strong differential rotation until then (e.g. Yoon \& Langer 2004).
We assume that the WD can no longer increase in mass when $|\dot{M}_{\rm 2}| < \dot{M}_{\rm r}$
for the differential rotation population (see also Chen \& Li 2009; Hachisu et al.\ 2012a,b).
Note that this assumption is only applied to the differential rotation
population (no critical accretion rate is in principle needed to sustain
solid-body rotation).

By using the Eggleton stellar evolution code (Eggleton 1973), we have calculated the evolution of the WD + MS systems.
The input physics for this code was updated over the past four decades (Han, Podsiadlowski \& Eggleton 1994; Pols et al.\ 1995, 1998;
Nelson \& Eggleton 2001; Eggleton \& Kiseleva-Eggleton 2002). The description of Roche-lobe overflow from Han, Tout \& Eggleton (2000) is adopted. The ratio of mixing length to local pressure scale height is set to be
2.0, and the convective overshooting parameter, $\delta_{\rm OV}$, to be 0.12, which roughly corresponds to an overshooting length of $\sim$0.25 pressure scale heights (e.g. Pols et al.\ 1997). A typical Population I composition is used, i.e. metallicity $Z=0.02$, H
abundance $X=0.70$ and He abundance $Y=0.28$.

The prescription of Hachisu et al.\ (1999) is adopted for the mass growth of the WD  by accretion of
H-rich matter from its mass-donating star (for details see Han \& Podsiadlowski 2004; Wang, Li \& Han 2010).
If the mass-transfer rate $|\dot{M}_{2}|$ is above a critical rate, $\dot{M}_{\rm cr}$, we assume that the accreted H steadily burns
on the surface of the WD with a rate of $\dot{M}_{\rm cr}$, and that the unprocessed matter is lost from the system as an
optically thick wind at a mass-loss rate $\dot{M}_{\rm wind}=|\dot{M}_{\rm 2}|-\dot{M}_{\rm cr}$ (Hachisu, Kato \& Nomoto 1996).
The critical mass-transfer rate is
 \begin{equation}
 \dot{M}_{\rm cr}=5.3\times 10^{\rm -7}\frac{(1.7-X)}{X}(M_{\rm
 WD}/\rm M_{\odot}-0.4) \,\rm M_{\odot}\,yr^{-1},
  \end{equation}
in which $X$ is the H mass fraction and $M_{\rm
 WD}$ is the mass of the WD.

We adopt the following assumptions when $|\dot{M}_{2}|$ is smaller than $\dot{M}_{\rm cr}$.
(1) When $|\dot{M}_{2}|$ is less than $\dot{M}_{\rm cr}$ but higher than
$\dot{M}_{\rm st}=\frac{1}{2}\dot{M}_{\rm cr}$, the H-shell burning is steady and no mass is lost
from the system. (2) When $|\dot{M}_{2}|$ is lower than $\dot{M}_{\rm st}$ but higher than
$\dot{M}_{\rm low}=\frac{1}{8}\dot{M}_{\rm cr}$, a very weak H-shell flash is triggered but
no mass is lost from the system. (3) When $|\dot{M}_{2}|$ is lower than $\dot{M}_{\rm low}$,
the H-shell flash is so strong that no material is accumulated onto the surface of the WD. We define
the mass-growth rate of the He layer under the H-shell burning as
 \begin{equation}
 \dot{M}_{\rm He}=\eta _{\rm H}|\dot{M}_{2}|,
  \end{equation}
in which $\eta _{\rm H}$ is the mass-accumulation efficiency for H-shell burning.
The values of $\eta _{\rm H}$ are:
 \begin{equation}
\eta _{\rm H}=\left\{
 \begin{array}{ll}
 \dot{M}_{\rm cr}/|\dot{M}_{2}|, & |\dot{M}_{2}|> \dot{M}_{\rm
 cr},\\
 1, & \dot{M}_{\rm cr}\geq |\dot{M}_{2}|\geq\dot{M}_{\rm low},\\
 0, & |\dot{M}_{2}|< \dot{M}_{\rm low}.
\end{array}\right.
\end{equation}
He is assumed to be ignited when the mass of the He layer reaches a certain value.

The WD mass-growth rate, $\dot{M}_{\rm CO}$, is defined as
 \begin{equation}
 \dot{M}_{\rm CO}=\eta_{\rm He}\dot{M}_{\rm He}=\eta_{\rm He}\eta_{\rm
 H}|\dot{M}_{\rm 2}|,
  \end{equation}
in which $\eta_{\rm He}$ is the mass-accumulation efficiency for He-shell burning.
The values of $\eta_{\rm He}$ are linearly interpolated from a grid, in which a wide range of WD masses and
mass-transfer rates were calculated in the He-shell flashes
(see Kato \& Hachisu 2004). Note that the accumulation efficiencies used
in this paper were derived for non-rotating WD
models, but it is assumed that they are still valid for rotating WDs.\footnote{It has been suggested that
rotation plays an important role on the accumulation efficiencies (see Piersanti et al. 2003).
Especially, He burning has been found to be  much less
violent when rotation is taken into account (e.g. Yoon, Langer \& Scheithauer 2004),
which may significantly increase the He accretion efficiency.}

The prescriptions above are incorporated into the Eggleton stellar
evolution code, and the evolution of the WD + MS systems is followed.
The mass lost from these WD binaries is assumed to carry away the
specific orbital angular momentum of the WD, whereas the mass loss induced by the donor's
wind is supposed to be negligible (e.g. Wang, Li \& Han 2010). Finally,
a large and dense model grids are obtained.

\section{Results of binary evolution calculations}

\begin{figure*}
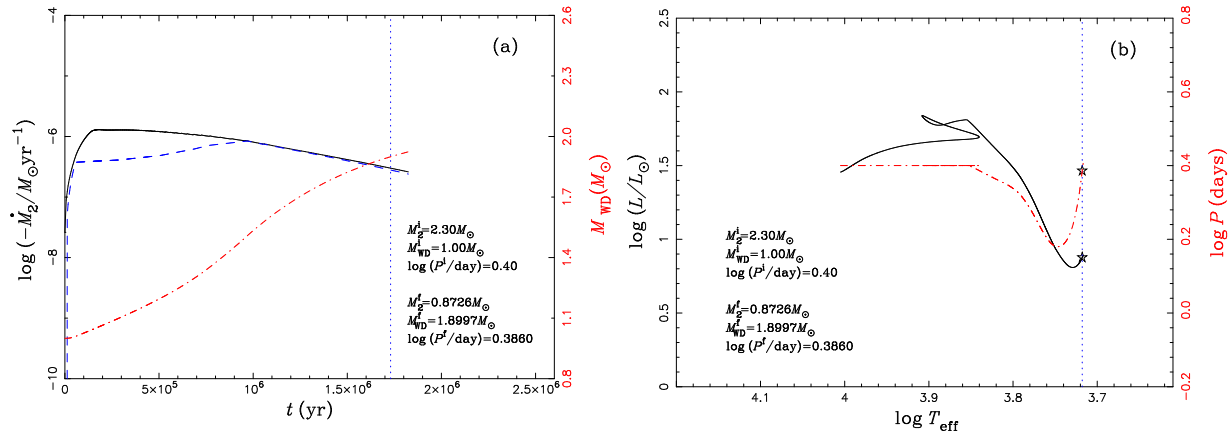

\centerline{\epsfig{file=f1a.ps,angle=270,width=8cm}\ \
\epsfig{file=f1b.ps,angle=270,width=8cm}} \caption{A representative example of binary evolution calculations.
Panel (a): the evolution of the mass-transfer rate (solid curve), the mass-growth rate of
the WD (dashed curve) and the mass of the WD (dash-dotted curve) as a function of time for the binary calculation.
Panel (b): luminosity of the mass-donating star (solid curve, left-hand axis) and
binary orbital period (dash-dotted curve, right-hand axis) as a function of effective temperature. Dotted vertical lines in both
panels and asterisks in panel (b) indicate the position where the WD increases the mass to its maximum.
The initial binary parameters and the parameters when the WD mass grows to its maximum are also given in
these two panels.}
\end{figure*}

In Fig.\,1, we present a representative case of the binary evolution calculations.
The WD explosion mass in this case is higher than
$1.5\,\rm M_{\odot}$, i.e.\ a case in which the WD requires differential rotation to postpone
explosion or collapse. The initial binary parameters are ($M_2^{\rm i}$, $M_{\rm WD}^{\rm i}$, $\log (P^{\rm i}/{\rm day})$)
$=$ (2.30, 1.00, 0.40), in which $M_2^{\rm i}$, $M_{\rm WD}^{\rm i}$ and $P^{\rm i}$ are the initial
mass of the MS star and of the WD in solar masses, and the initial orbital period in days,
respectively.  In this case, the mass-donating star fills its Roche lobe in the subgiant stage, which results
in early Case B mass transfer. The mass-transfer rate $|\dot{M}_{\rm 2}|$ exceeds $\dot M_{\rm cr}$
soon after the start of Roche-lobe overflow, leading to a wind phase in which a part of the transferred mass
is blown off in the form of the optically thick wind, and the rest is accumulated onto the surface
of the WD. After about $10^{6}$\,yr, $|\dot{M}_{\rm 2}|$ drops below $\dot M_{\rm cr}$ but remains
higher than $\dot M_{\rm st}$. Thus, the optically thick wind stops and the H-shell burning is
stable. With the continuing decrease of $|\dot{M}_{\rm 2}|$, the system enters into
a weak H-shell flash phase. The WD always grows in mass until the mass-transfer rate decreases
to $\dot{M}_{\rm r}$.  At this moment, the mass of the WD is $M_{\rm WD}^{\rm f}=1.8997\,\rm M_{\odot}$,
the mass of the donor star is $M^{\rm f}_2=0.8726\,\rm M_{\odot}$, and the orbital period
$\log (P^{\rm f}/{\rm day})=0.3860$.

\begin{figure*}
\begin{center}
\epsfig{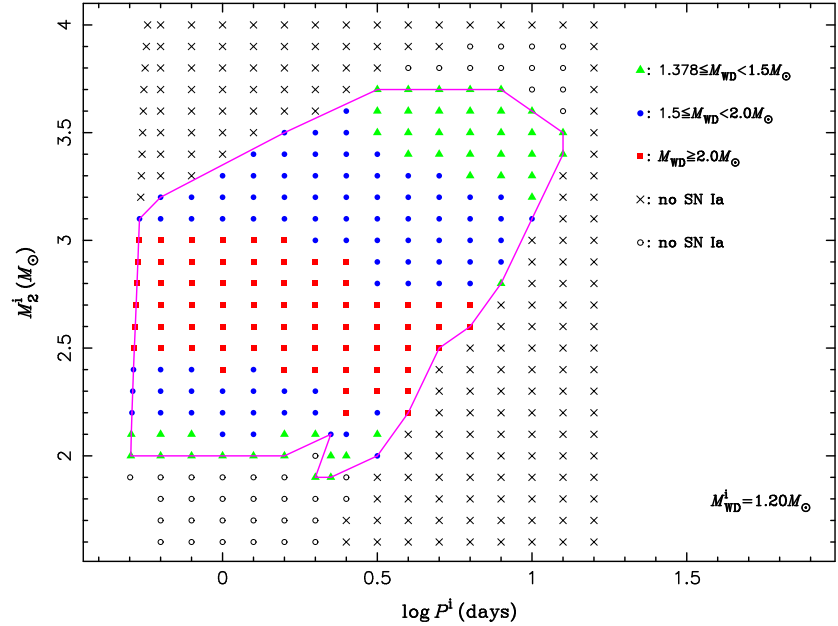}
\caption{Final results of binary evolution calculations
in the initial orbital period--secondary mass ($\log P^{\rm i}$, $M^{\rm i}_2$) plane of
the WD + MS system for an initial WD mass of 1.2$\,\rm M_{\odot}$. The filled symbols represent
systems resulting in \textbf{a} SN Ia explosion. The green triangles, blue circles and red squares
denote WD explosion masses in the range of $1.378\leq M_{\rm WD}<1.5\,\rm M_{\odot}$, $1.5\leq M_{\rm WD}<2.0\,\rm M_{\odot}$,
and $M_{\rm WD} \geq 2.0\,\rm M_{\odot}$, respectively.
Open circles indicate systems that experience novae
which prevent the WD from reaching 1.378$\,\rm M_{\odot}$, whereas crosses represent those which experience
dynamically unstable mass transfer.}
\end{center}
\end{figure*}

\begin{figure*}
\begin{center}
\epsfig{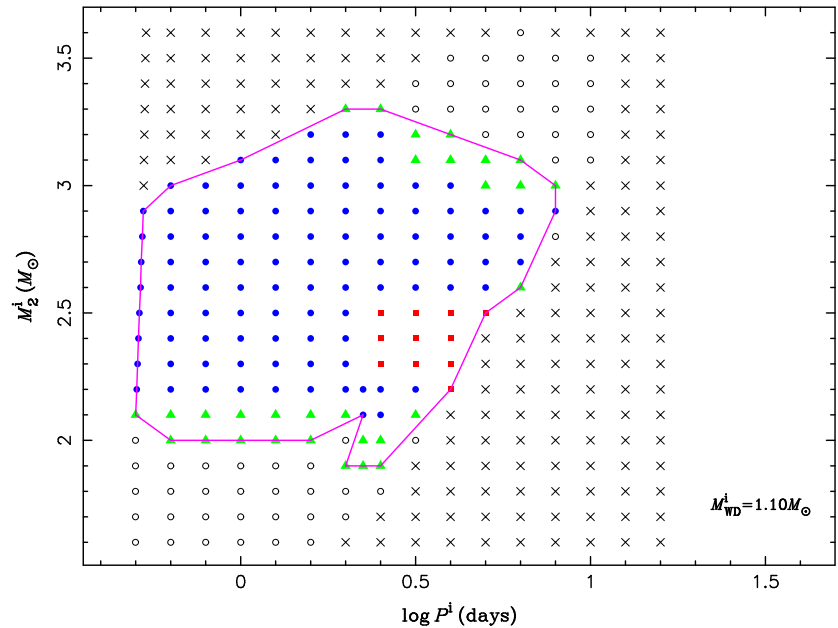}
\caption{As Fig.\,2, but for an initial WD mass
of 1.1$\,\rm M_{\odot}$.}
\end{center}
\end{figure*}

\begin{figure*}
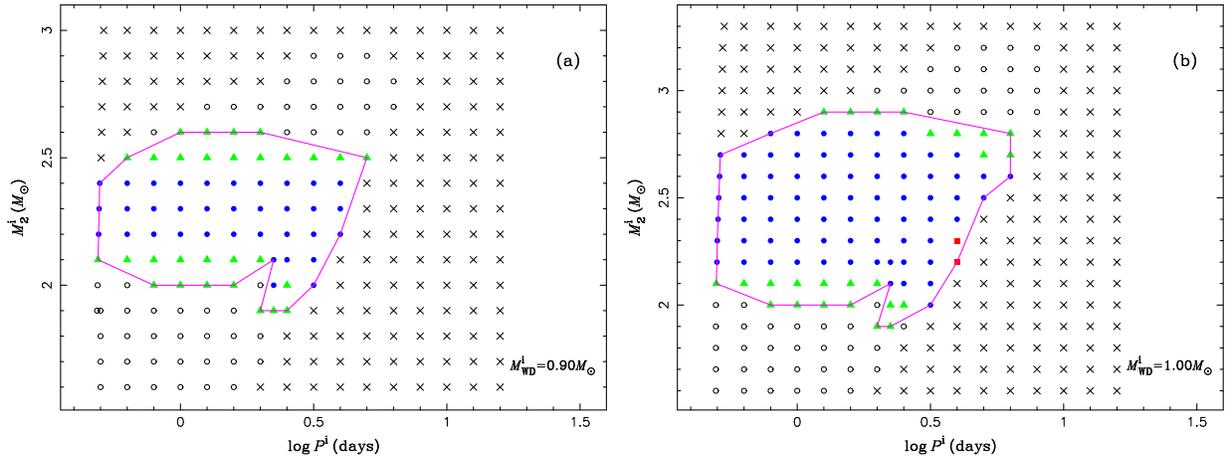

\begin{center}
\centerline{\epsfig{file=f4a.ps,angle=270,width=8cm}\ \
\epsfig{file=f4b.ps,angle=270,width=8cm}} \caption{As Fig.\,2, but for initial WD masses
of 0.9 and 1.0$\,\rm M_{\odot}$.}
\end{center}
\end{figure*}

\begin{figure*}
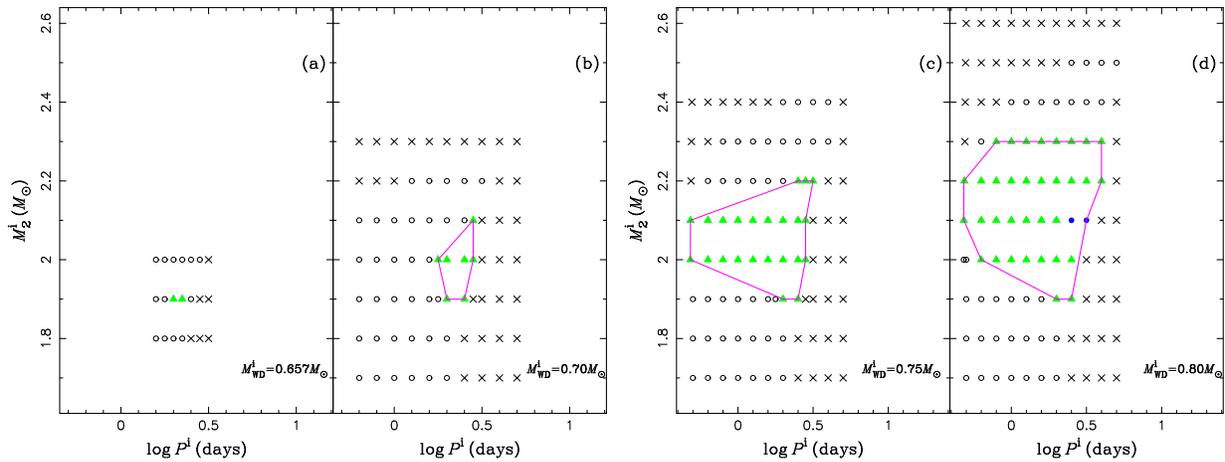

\begin{center}
\centerline{\epsfig{file=f5ab.ps,angle=270,width=8cm}\ \
\epsfig{file=f5cd.ps,angle=270,width=8cm}} \caption{As Fig.\,2, but for initial WD masses
of 0.657, 0.70, 0.75 and 0.80$\,\rm M_{\odot}$.}
\end{center}
\end{figure*}

%tentatively

Figs\,2--5 show the final outcomes of our binary evolution calculations in the initial orbital
period and secondary mass plane. The green triangles, blue circles
and red squares denote that the WD explodes as a SN Ia with the WD mass in the range of
$1.378\leq M_{\rm WD}<1.5\,\rm M_{\odot}$, $1.5\leq M_{\rm WD}<2.0\,\rm M_{\odot}$ and
$M_{\rm WD} \geq 2.0\,\rm M_{\odot}$, respectively. It is tempting to link WD explosion
mass ranges with observed SN Ia subtypes. For example, ``normal'' SNe Ia might naturally be
associated with the WD mass in the range of $1.378\leq M_{\rm WD}<1.5\,\rm M_{\odot}$, which can be
supported by solid-body rotation. The mass ranges displayed in our figures corresponding to the
divisions chosen by Hachisu et al.\ (2012b). They associated the middle mass range
($1.5\leq M_{\rm WD}<2.0\,\rm M_{\odot}$) with the brighter SN 1991T-like class, with the
acknowledged super-Chandrasekhar events from yet more massive WDs ($M_{\rm WD} \geq
2.0\,\rm M_{\odot}$). However, we note that Mazzali et al.\ (2007) inferred a mass for the progenitor
of a SN~1991T-like object which is consistent with the Chandrasekhar mass.
For a discussion about the ranges of WD explosion mass see Section 6.1.

Some simulated WD + MS binaries fail to produce
SNe~Ia due to nova explosions that prevents the WD growing in mass to 1.378$\,\rm M_{\odot}$
(open circles in Figs\,2--5), or dynamically unstable mass transfer
(crosses in Figs\,2--5). The maximum explosion mass of the WD in our calculations is $2.4558\,\rm M_{\odot}$.
We note that WDs supported by differential rotation might explode as SNe Ia soon after the WD mass exceeds
2.4$\,\rm M_{\odot}$ due to a secular instability at $T/|W|\sim0.14$, in which $T$ and $W$ are the rotational
and gravitational energies of the WD, respectively (e.g. Yoon \& Langer 2004; Hachisu et al.\ 2012a).  At this point,
the WD would be too massive to maintain hydrostatic equilibrium and would contract on a short timescale
due to angular momentum redistribution. As a result of such rapid compression, carbon seems likely to
be ignited in the central region of the WD, and then a SN~Ia explosion might occur. Our binary
evolution models predict that SNe Ia from this case are rare, since it only happens when
$M_{\rm WD}^{\rm i}\geq1.2\,\rm M_{\odot}$.

\section{Method of Binary population synthesis}

In order to obtain SN~Ia birthrates and delay times, and the properties of the mass-donating star
at the point when the WD increases the mass to its maximum, a series of Monte
Carlo simulations in the BPS approach are performed.  The following
initial conditions for the Monte Carlo simulations are adopted:
(1) The initial mass function of the primary star is from Miller \& Scalo (1979).
The mass of the primary star is from $0.1\,\rm M_{\odot}$ to $100\,\rm M_{\odot}$.
(2) A constant mass-ratio distribution is taken (e.g. Goldberg \& Mazeh 1994).
(3) The distribution of initial orbital separations is assumed to be constant in $\log a$ for wide
binary systems, where $a$ is orbital separation (e.g. Han, Podsiadlowski \& Eggleton 1995).
(4) A circular orbit is assumed for all binary systems.
(5) A constant star formation rate (SFR) is simply assumed over the past 14\,Gyr or,
alternatively, as a delta function, i.e. a single instantaneous starburst
(a burst producing $10^{10}\,\rm M_{\odot}$ in stars is assumed).
We intend the constant SFR to provide a rough description of spiral galaxies, and the delta
function to approximate elliptical galaxies (or star clusters, for which the normalization
would need to be altered).

For each BPS simulation, we employed
the Hurley binary evolution code  (Hurley, Pols \& Tout 2002) to
evolve $10^{\rm 7}$ sample binaries. These binaries are followed from
the star formation to the formation of the CO WD + MS systems based on
three binary evolutionary scenarios (i.e.\ \textit{He star}, \textit{E-AGB} and \textit{TP-AGB
scenarios}; for details see Section 4.2 of Wang, Li \& Han 2010).
The metallicity in these simulations is set to be $0.02$. If the initial
parameters of a WD + MS system at the start of the Roche-lobe overflow are
located in the SN Ia production regions in the plane of
($\log P^{\rm i}$, $M_2^{\rm i}$) for its specific $M_{\rm WD}^{\rm i}$,
a SN~Ia is assumed to occur, and the properties of the WD + MS system when the WD increases the mass to
its maximum are obtained by interpolation in the three-dimensional grid ($M_{\rm WD}^{\rm i}$,
$M_2^{\rm i}$, $\log P^{\rm i}$) of the WD + MS systems calculated in Section 3.

In the binary evolution, the WD + MS system is most likely produced from the
common-envelope evolution of a giant binary system.
Common-envelope evolution is still very poorly-understood
(e.g.\ Ivanova et al.\ 2013; Zuo \& Li 2014), so we use
the standard energy equations to estimate the outcome of each common-envelope phase (e.g.\ Webbink
1984). In this parametrization of common-envelope evolution,
there are two unknown parameters, i.e.\ $\alpha_{\rm ce}$ and $\lambda$, in which
$\alpha_{\rm ce}$ is the ejection efficiency of common-envelope energy and $\lambda$ is
a stellar structure parameter that relates to the definition of the core-envelope boundary and the evolutionary stage of the
mass-donating star. Similar to our previous works (e.g.\ Wang et al. 2009b),
we used a single free parameter $\alpha_{\rm ce}\lambda$ to describe the process of the common-envelope ejection, and gave
the results for two specific values (i.e.\ 0.5 and 1.5).

\section{Results of binary population synthesis}

\subsection{Birthrates and delay times of SNe Ia}

\begin{figure}
\begin{center}
\includegraphics[width=6cm,angle=270]{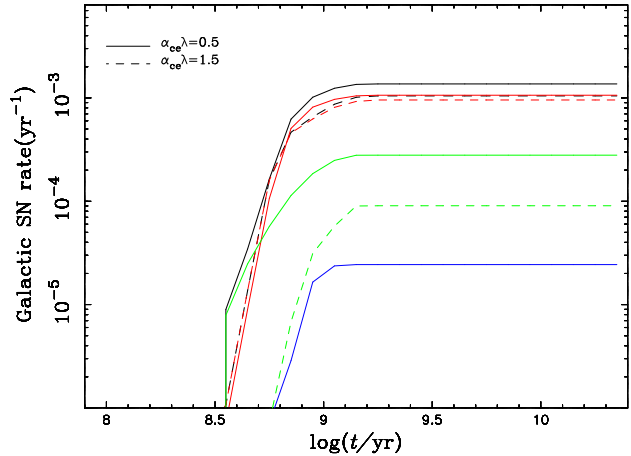}
 \caption{Evolution of Galactic SN Ia birthrates with time for a constant Population I SFR ($Z=0.02$,
${\rm SFR}=5\,\rm M_{\odot}{\rm yr}^{-1}$). The black curves are for the total
SN Ia birthrates. The red, green and blue curves are for SNe Ia with WD explosion
masses in the range of 1.378--1.5$\,\rm M_{\odot}$, 1.5--2$\,\rm M_{\odot}$ and
$\geq$$2\,\rm M_{\odot}$, respectively. The solid and dashed curves show the results
of different common-envelope ejection parameters with $\alpha_{\rm ce}\lambda=0.5$ (solid) and
$\alpha_{\rm ce}\lambda=1.5$ (dashed), respectively. Note that there is no blue dashed
curve in this figure, for the reason see the text.}
  \end{center}
\end{figure}

The observed SN~Ia birthrate in our Galaxy is $\sim$$3-4\times10^{-3}\,{\rm yr}^{-1}$
(e.g. Cappellaro \& Turatto 1997), which can be used to constrain the progenitor models
of SNe~Ia. In Fig.\,6, we show the evolution of Galactic SN Ia birthrates from different
ranges of WD explosion mass by adopting $Z=0.02$ and ${\rm SFR}=5\,\rm M_{\odot}{\rm yr}^{-1}$.
According to our standard model ($\alpha_{\rm ce}\lambda=0.5$) for the WD + MS channel,
the simulation gives the Galactic total SN Ia birthrate of $\sim$$1.36\times 10^{-3}\ {\rm yr}^{-1}$
(black solid curve in Fig.\,6). However, the birthrate from $\alpha_{\rm ce}\lambda=1.5$
is lower than that of $\alpha_{\rm ce}\lambda=0.5$; the simulation with $\alpha_{\rm ce}\lambda=1.5$
gives a total SN Ia birthrate of $\sim$$1.05\times 10^{-3}\ {\rm yr}^{-1}$ (black dashed curve
in Fig.\,6). This is because the binaries which result from common-envelope ejections tend to have
slightly closer orbits for $\alpha_{\rm ce}\lambda=0.5$ and are
therefore more likely to be located in the
SN Ia production region. From this simulation, the birthrate of SNe Ia with WD explosion masses $\geq$$2\,\rm M_{\odot}$
is $\sim$$2.44\times 10^{-5}\ {\rm yr}^{-1}$ based on our standard model, whereas the simulation with $\alpha_{\rm ce}\lambda=1.5$
does not produce these massive WDs. This is due to the lack of WDs with initial masses
$\geq$$1\,\rm M_{\odot}$ for the simulation of $\alpha_{\rm ce}\lambda=1.5$ (see Fig.\,9).

\begin{figure}
\begin{center}
\includegraphics[width=6cm,angle=270]{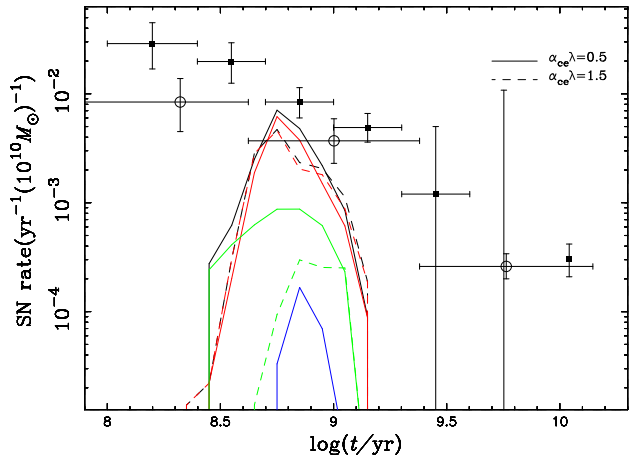}
 \caption{As Fig.\,6, but for a single
starburst with a total mass of $10^{10}\rm M_{\odot}$.  The open circles are from Maoz et al. (2011), and the filled squares from Totani et al. (2008).  }
  \end{center}
\end{figure}

The theoretical delay time distributions of SNe Ia can be compared with that of
observations, and then used to examine current progenitor models (e.g. Mennekens et al. 2010; Meng \& Yang 2010).
In Fig.\,7, we present the evolution of SN Ia birthrates for a single
starburst with a total mass of $10^{10}\,M_{\odot}$.
From this figure, we can see that a high value of $\alpha_{\rm ce}\lambda$ leads to
a systematically later explosion time for the WD + MS channel. This is
because a high value of $\alpha_{\rm ce}\lambda$ leads to wider WD binaries, and, as a consequence,
it takes a longer time for the companion to fill its Roche lobe. This figure also shows
that SN explosions from the WD + MS channel have delay times of $\sim$250\,Myr$-$1.4\,Gyr,
which suggests that this channel only has a contribution to part of the overall SNe Ia.
For other potential SN Ia production methods see  Wang, Justham \& Han(2013),
Soker, Garc\'{\i}a-Berro \& Althaus (2014) and Meng \& Podsiadlowski (2014).

Compared to some recent studies (e.g.
Ruiter, Belczynski \& Fryer 2009; Bours, Toonen \& Nelemans 2013),
we have obtained higher SN Ia birthrates for the WD + MS channel.
The main difference is that they adopted the efficiency of mass
accumulation on the surface of WDs from Prialnik \& Kovetz (1995),
which is significantly lower than that assumed in this work.
However, the specific efficiency of mass accumulation is still
uncertain (see Cassisi, Iben \& Tornamb\`{e} 1998).

After the WD accretes matter to reach the final explosion mass, the WD probably needs a spin-down time
before it explodes (e.g. Justham 2011; Di Stefano, Voss \& Claeys 2011; Hachisu et al. 2012a,b).
The time delay after the end of the mass transfer whilst the WD internally redistributes or loses spin
angular momentum -- ``spins down'' -- may be the most natural way by which the SD model can satisfy
observational constraints which would otherwise be problematic for it.
However, the spin-down time is still uncertain.
(1) For WDs which can be supported by rigid-body rotation, the spin-down time is mainly determined by
angular momentum loss from the WD. In the absence of other braking mechanisms, this likely depends
on the strength of the WD magnetic field, e.g. spin down may take more than $10^9$\,yr for
magnetic fields of $\sim$$10^6$\,G (e.g. Ilkov \& Soker 2011).
(2) For WDs massive enough to require the
support of differential rotation then, as angular momentum in the WD core is lost or redistributed
toward rigid-body rotation, the WD core will contract until its central density and temperature become
high enough to ignite carbon.
Meng \& Podsiadlowski (2013) recently argued that the upper
limit of the spin-down timescale is a few $10^7$\,yr for SN Ia progenitor systems that contain a red giant
donor. Note that the effect of spin-down time on the delay time distributions of SNe Ia is neglected in Fig.\,7 due to
the uncertainties of spin-down timescale.

\subsection{Distribution of WD explosion masses}

\begin{figure}
\begin{center}
\includegraphics[width=10cm,angle=0]{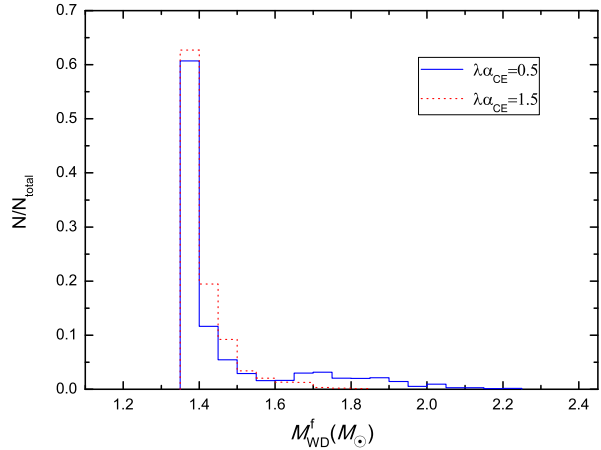}
 \caption{Distribution of WD explosion masses for the WD + MS channel with different values of $\alpha_{\rm ce}\lambda$.}
  \end{center}
\end{figure}

Fig.\,8 shows the distribution of the WD explosion mass with different assumed values of $\alpha_{\rm ce}\lambda$.
The simulation predicts a range of WD explosion masses extending from 1.378 to $>$$2\,\rm M_{\odot}$.
With $\alpha_{\rm ce}\lambda=0.5$, 77\% of these exploding WDs are predicted to have masses in
the range of 1.378$-$1.5$\,\rm M_{\odot}$, i.e.\,to be WDs which can be supported by rigid-body rotation,
23\% have progenitor masses $>$1.5$\,\rm M_{\odot}$ which require differential rotation for support
(only 2\% of the total in this study have WD explosion masses $\geq$2.0$\,\rm M_{\odot}$; these SNe require the
initial mass of the WD to be larger than 1.0$\,\rm M_{\odot}$). Over-luminous events clearly form
a relatively rare subclass of SNe Ia, and they have WD explosion mass more massive than $2.0\,\rm M_{\odot}$;
about 1\% SNe Ia can be of this type (e.g. Howell et al. 2006;  Hicken et al. 2007; Scalzo et al.
2010; Silverman et al. 2011). If the WD + MS channel has a contribution to the total SNe Ia in
the observations, the simulation from this work will occupy $\sim$2\% over-luminous events based
on our standard model, which is comparable with that of observations.
Additionally, a higher proportion of the SN Ia population comes from WDs which could
be supported by rigid-body rotation until explosion when compared with
previous work (for a discussion of this see Section 6.1).

\subsection{Initial parameters of WD + MS systems}

\begin{figure}
\begin{center}
\includegraphics[width=10cm,angle=0]{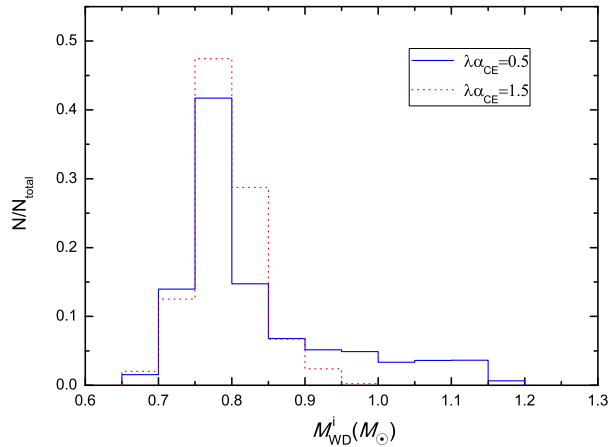}
 \caption{Distribution of the initial WD masses for WD + MS systems that ultimately
 produce SNe Ia with different values of $\alpha_{\rm ce}\lambda$.}
  \end{center}
\end{figure}

\begin{figure}
\begin{center}
\includegraphics[width=10cm,angle=0]{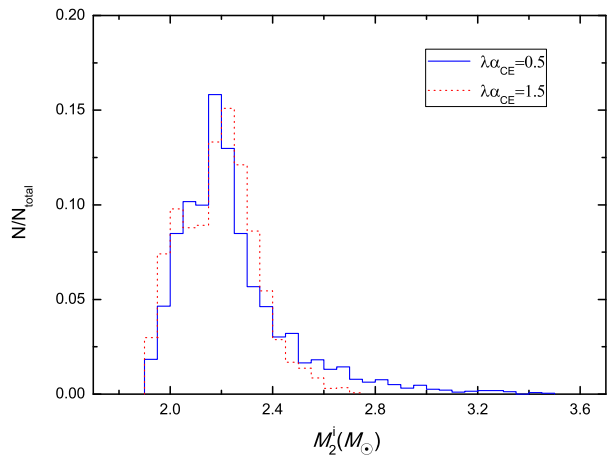}
 \caption{As Fig.\,9, but for the distribution of the initial masses of secondaries in
 WD + MS systems.}
  \end{center}
\end{figure}

\begin{figure}
\begin{center}
\includegraphics[width=10cm,angle=0]{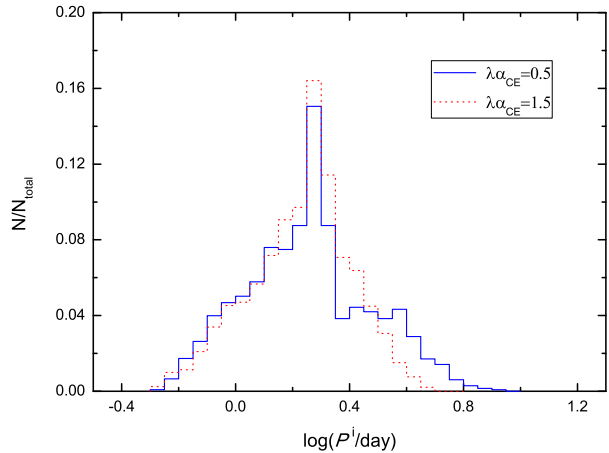}
 \caption{As Fig.\,9, but for the distribution of the initial orbital periods of WD +
 MS systems.}
  \end{center}
\end{figure}

Some WD + MS systems are candidates to be SN Ia progenitors in the observations, for
reviews see Wang \& Han (2012) and Parthasarathy et al.\ (2007). According to
our BPS calculations, we can present some properties of initial WD + MS systems for
producing SNe Ia, which would be helpful to search potential progenitor candidates of
SNe Ia.

Fig.\,9 shows the distribution of initial WD masses in WD + MS systems that ultimately
produce SNe Ia for different values of $\alpha_{\rm ce}\lambda$. The solid and dotted
histograms represent the cases with $\alpha_{\rm ce}\lambda=0.5$ and $\alpha_{\rm ce}\lambda=1.5$,
respectively. The simulation uses a metallicity of $Z=0.02$ and a constant initial mass-ratio
distribution. This figure displays a result of the current epoch for a constant SFR. From this
figure, we can see that the distribution for $\alpha_{\rm ce}\lambda=0.5$ has a high-mass tail,
but this does not appear in the case of $\alpha_{\rm ce}\lambda=1.5$. These massive CO WDs are
mainly from \textit{TP-AGB scenario} (see Section 4.2 of Wang, Li \& Han 2010).
Because of the low binding energy of the common envelope
and the long primordial orbital period in the \textit{TP-AGB scenario}, $\alpha_{\rm ce}\lambda$
has a significant influence on the formation of the WD + MS systems; if a common envelope can be ejected,
a low $\alpha_{\rm ce}\lambda$ value produces a WD + MS system with shorter orbital period
that is more likely to fulfil the conditions for producing SNe Ia. Thus, we can see obvious
contributions from  the \textit{TP-AGB scenario} when $\alpha_{\rm ce}\lambda=0.5$, but very
little contribution when $\alpha_{\rm ce}\lambda=1.5$.

Fig.\,10 represents the distribution of the initial masses of secondaries in WD + MS
systems for producing SNe Ia. From this figure, we can see that the distribution for
$\alpha_{\rm ce}\lambda=0.5$ has a high-mass tail that is from  the contribution of the
\textit{TP-AGB scenario}. In addition, a more massive secondary in a WD + MS system will evolve
more quickly and thus produce a SN Ia at an earlier time. In Fig.\,11, we display the
distribution of the initial orbital periods of WD + MS systems for the production of SNe Ia.
Similar to the distributions of the initial masses of CO WDs and those of the initial
orbital periods, the long orbital periods for $\alpha_{\rm ce}\lambda=0.5$ are also
from the \textit{TP-AGB scenario}.

\section{Discussion} \label{6. DISCUSSION}

\subsection{WD explosion mass}
The WDs with masses that could be supported by rigid-body rotation
until explosion from
Hachisu et al. (2012b) account for 48\% of the total.\footnote{Hachisu et al. (2012b)
did not employ detailed BPS calculations to obtain this value.}
However, the result from our calculations is 77\%.
The main difference is that they did not employ full stellar evolution calculations;
they adopted a simple analytical fitting formulae for estimating the mass-transfer rate
(for details see Section 5.1 of Han \& Podsiadlowski 2004).
Specifically, the minimum initial WD mass for producing SNe Ia from
Hachisu et al. (2012b) is about 0.7$\,\rm M_{\odot}$, but only few SNe Ia are obtained from this WD
mass due to very small region for producing SNe Ia.
However, we obtained more SNe Ia for low-mass initial WDs, which have a higher contribution to SN Ia population
that could be supported by rigid-body rotation.
Our substantially higher fraction of WDs with
explosion masses of 1.5$\,\rm M_{\odot}$ or less should be helpful with
reconciling these SD SNe Ia with the inference that most SNe Ia have
masses close to the Chandrasekhar mass (Mazzali et al.\ 2007).

Hachisu et al.\ (2012b) have chosen to map the observed subtypes of SNe Ia onto the full range of
the WD explosion masses. They associated the rigid-body rotation population (1.378$-$1.5$\,\rm M_{\odot}$)
with ``normal'' SNe Ia and 91bg-like events, and the differential rotation population ($>$$1.5\,\rm M_{\odot}$) with the brighter
91T-like class and over-luminous events.
However, the work by Mazzali et al. (2007) concludes that the pre-explosion WD
masses of a diverse sample of SNe Ia is consistent with the Chandrasekhar mass, in which the sample
includes 91bg-like, spectroscopically normal and 91T-like events. Without a systematic bias
in either method or sample of events, their results suggest that it is unlikely that the
class of 91T-like SNe includes events with pre-explosion masses as high as 2$\,\rm M_{\odot}$, as
adopted by Hachisu et al.\ (2012b). Even so, we note that an updated analysis of 91bg-like events
by Mazzali \& Hachinger (2012) does allow the progenitor masses for this class of SNe Ia to be
lower than the masses indicated by the study of Mazzali et al. (2007). Interestingly, the work by Mazzali
et al.\ (2007) indicates that the normal range of SNe Ia arises from WD masses which are low
enough to be supported by near solid-body rotation (i.e.\,within $\approx$$0.1\,\rm M_{\odot}$ of
the normally-assumed ignition mass, 1.378$\,\rm M_{\odot}$).

Justham (2011) suggested that future BPS calculations should investigate to what extent typical
WD masses in SD SNe Ia would exceed the Chandrasekhar mass if the spin of the accreting WD delays
the explosion. This work has investigated that question. For the standard assumptions in our
model, most WDs (77\%) have low enough masses
to be supported by solid-body rotation. This helps to support the spin-up/spin-down scenario as
a viable explanation for the lack of H in SNe Ia if they are produced from SD progenitor
systems.

\subsection{Post-accretion system appearance and post-explosion remnant properties}

One of the most exciting possibilities arising from including the spin-down time of WDs in SD
SN Ia models is that we would, in principle, be able to observe systems in which the WD
is already more massive than $M_{\rm Ch}$. In some cases, the WD would no longer be accreting;
it would be losing or internally redistributing angular momentum whilst waiting to explode.
We may have already known such systems. For example, in U Scorpii, the uncertainty in the dynamical
mass measurement would allow the WD to  be more massive than $M_{\rm Ch}$ (e.g. Thoroughgood
et al.\ 2001). The novae RS Ophiuchi and V445 Puppis are each also inferred to contain a WD
with a mass very close to the Chandrasekhar limit, see Sokoloski et al.\ (2006) and Kato et al.\
(2008), respectively.

\begin{figure}
\begin{center}
\includegraphics[width=7cm,angle=270]{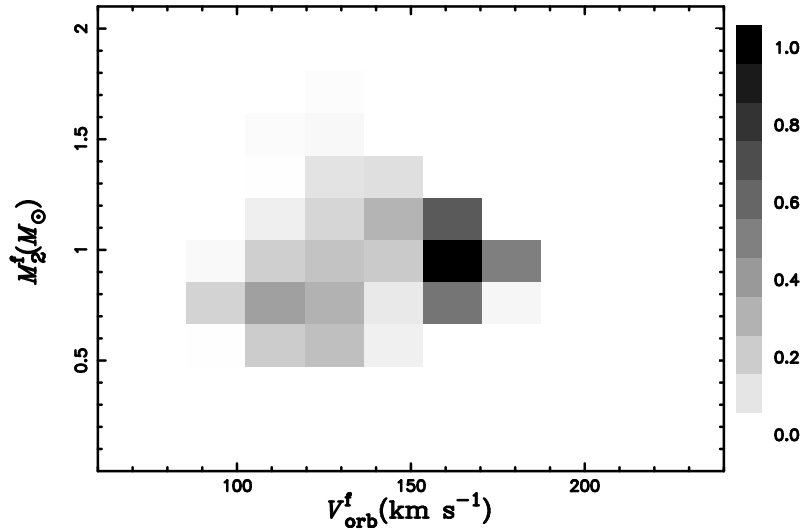}
 \caption{The distribution of properties of the companions in the plane of ($V_{\rm orb}^{\rm f}$,
 $M_2^{\rm f}$) at the current epoch, in which $V_{\rm orb}^{\rm f}$ is the orbital velocity and $M_2^{\rm f}$
 the companion mass when the WD mass reaches its maximum. Here, we set $\alpha_{\rm ce}\lambda=0.5$.}
\end{center}
\end{figure}

The mass-donating star in the SD model would survive after the SN explosion and potentially
be identifiable soon after the WD is disrupted (e.g. Justham et al. 2009; Wang \& Han 2009, 2010;
Liu et al. 2012, 2013; Pan, Ricker \& Taam 2014). The BPS gives current-epoch distributions of many properties of companions when the WD
increases the mass to its maximum (e.g. the orbital velocities,
the luminosities, the surface gravities, the effective temperatures, the
surface abundances, etc). These properties may be a starting point to study the surviving companion stars of
SNe Ia.

Fig. 12 shows an example of the distributions of the properties of companion stars in
the orbital velocity and companion mass plane at the point when the WD mass has increased to its
maximum, which may be helpful for identifying the surviving companions.
The star known as Tycho G was suggested to be a possible
surviving companion star from the system which produced Tycho's SN (a Galactic SN Ia) by Ruiz-Lapuente et al.\,(2004).
They found that this star have a spatial velocity of $136\,{\rm km/s}$  that is more than three times the mean velocity of the stars in the vicinity. This velocity is compatible with the value obtained from this study. However, whether Tycho G is
the surviving companion of Tycho's SN is still debatable
(see, e.g. Fuhrmann 2005; Kerzendorf et al. 2009, who found no
evidence that Tycho G has an anomalous space velocity).
Additionally, the final properties of the
surviving companions also depend on the spin-down time of WDs.
If the spin-down time is long enough (e.g. $\sim$$10^9$\,yr), the companion will
evolve to a He WD or CO WD  when SN explosion occurs.  In such
case, any prominent signature of the companion is not expected immediately
before or after the SN explosion (e.g. Justham 2011; Di Stefano, Voss \& Claeys 2011).

\subsection{SNe Ia with circumstellar material and apparent SNe IIn}

The population of SN Ia progenitors in which the WD must be supported by differential rotation
contains systems which might naturally produce a signature of H in SNe Ia.  The mass-donating stars in these systems
could still have substantial H envelope masses at SN explosion. In many cases, the WDs could
still be accreting at a high mass-transfer rate at the moment of the explosion -- or would have
very recently been doing so -- and would therefore be surrounded by dense, H-rich, circumstellar
material. These systems might therefore help to explain two different phenomena. Firstly, it is
now widely recognized that some SNe Ia (e.g.\,SN 2002ic-like objects) contain H at the moment of explosion
although other models exist to explain these events (see, e.g. Hamuy et al. 2003; Livio \& Riess 2003;
Chugai \& Yungelson 2004; Han \& Podsiadlowski 2006; Wood-Vasey \& Sokoloski 2006; Dilday et al. 2012).
Secondly, and less well known, observational evidence has been increasing which indicates that a large
fraction of the events classified as type IIn supernovae (SNe IIn) are not produced by massive stars
(e.g. Kotak et al. 2004; Anderson et al. 2012), and a natural candidate to explain some of these
events would be SNe Ia which explode inside a dense H-rich
environment.

It is therefore not a new idea that some SNe which exhibit SN IIn phenomenology
might be disguised SNe Ia. However, we suggest that many SNe which
exhibit SN IIn phenomenology might be produced by explosions of WDs
that had been supported by differential rotation, and that consequently the progenitor systems were
H-rich at the time of SN explosion.
If all WDs that could be supported by differential rotation have a contribution to SNe IIn,
the birthrates from this scenario are $\sim$$0.31\times 10^{-3}\,{\rm yr}^{-1}$ based on our standard model.
Unfortunately, the birthrates of SNe IIn are very uncertain, but some
observational estimates give $\sim$$0.13-1.22\times 10^{-3}\ {\rm yr}^{-1}$
(Cappellaro \& Turatto 1997; Smartt et al. 2009).
Therefore, this scenario might produce a significant part of the population of SNe
classified as type IIn. In this case then the WD masses of such
disguised SNe Ia would be higher than those of typical SNe Ia; if
those WD masses can be inferred, then it would provide a test of this model.
We note that there is no H in observed over-luminous events, so
the estimated birthrate for these SNe will become lower if the WD ($\geq$2.0$\,\rm M_{\odot}$)
is still accreting matter with a high mass-transfer rate at the moment of SN explosion.

\subsection{SN Ia diversity}

The overall origin of SN Ia systematic diversity is probably complex.
We stress that we are not trying to explain the differences
between individual SN Ia, but to explore reasons why different
stellar populations show systematic differences in the SNe Ia which
they produce. To first order, it is expected that the $^{56}$Ni mass
produced by each SN Ia controls the maximum luminosity (Arnett
1982), but the origin of the variation of the $^{56}$Ni mass for different
individual SN~Ia is still uncertain (one possibility is the location
of the deflagration-to-detonation transition; see also the
Introduction).\footnote{Metallicity may also affect the appearance of SNe~Ia (see
H\"{o}flich, Wheeler \& Thielemann 1998;  Dom\'{\i}nguez et al. 2001; Timmes, Brown \& Truran 2003; Podsiadlowski et al. 2006;
Bravo et al. 2010; Sullivan et al. 2010; Childress et al. 2013; Pan et al. 2014).}
Whilst it is expected to be possible to have a wide
variation in $^{56}$Ni mass produced by identical WD explosion masses,
it is obvious that differences in WD explosion mass may have an
influence on $^{56}$Ni masses in SNe Ia. If there were systematic population
variations in WD masses with age, this might help to explain recent
observations which find variations in SN Ia population properties with
the age of the host stellar population (see, e.g. Howell et al. 2009; Pan et al. 2014).

\begin{figure}
\begin{center}
\includegraphics[width=11cm,angle=0]{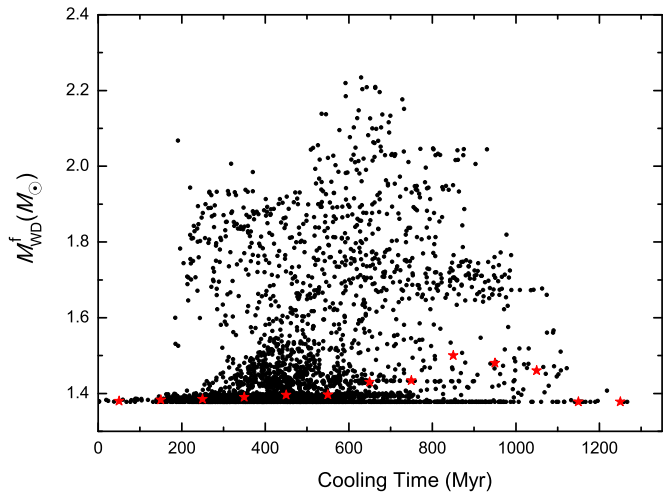}
 \caption{WD explosion mass versus WD cooling time before mass accretion, in which we set $\alpha_{\rm ce}\lambda=0.5$.
The black circles denote the WD explosion mass versus WD
  cooling time for individual instances produced by our Monte-Carlo
  population synthesis. The red stars present the median WD explosion
  mass versus WD cooling time; these indicate that the majority of
  the SNe have explosion masses close to the Chandrasekhar mass.}
  \end{center}
\end{figure}

\begin{figure}
\begin{center}
\includegraphics[width=11cm,angle=0]{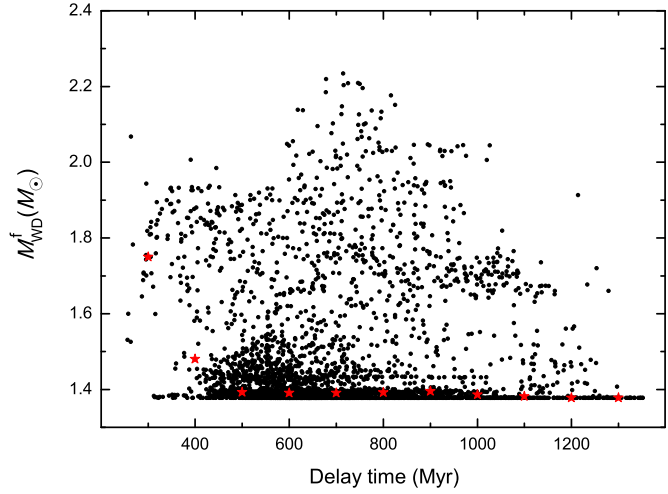}
\caption{As Fig. 13, but for WD explosion mass versus SN Ia delay
   time. We stress that these delay times do not include any
   contribution from spin-down times. Note that for a delay time
   of 300 Myr the median WD explosion
 mass is significantly above 1.5\,$\rm M_{\odot}$.}
  \end{center}
\end{figure}

\begin{figure}
\begin{center}
\includegraphics[width=11cm,angle=0]{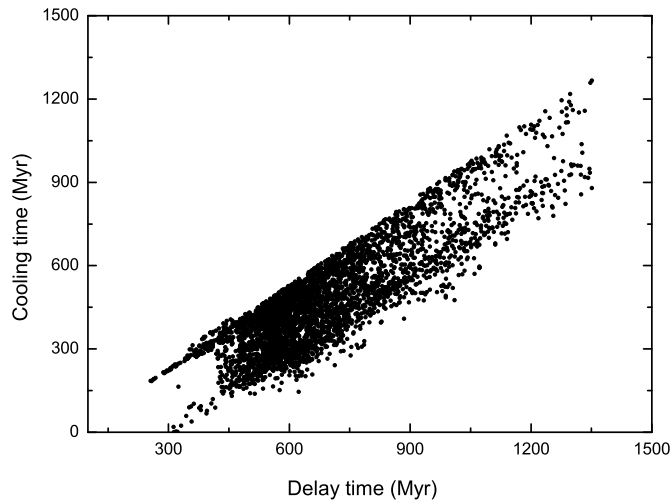}
\caption{WD cooling time before mass accretion versus the delay time of
SNe Ia, in which we set $\alpha_{\rm ce}\lambda=0.5$.}
\end{center}
\end{figure}

Not only should accounting for the accreting WD's spin angular momentum result
in a range of WD masses at explosion, but it should also lead to a range of
post-accretion cooling times before SN explosion due to the duration of
the phase of spin-down or internal angular momentum redistribution (see
also Hachisu et al. 2012a).
In addition to this post-accretion cooling time, the pre-accretion
cooling times also affect the material properties of the WD at explosion,
in particular the density at SN explosion that may be the origin of the
maximum luminosity scatter (e.g. Lesaffre et al.\ 2006; Krueger et
al. 2010). WDs tend to crystallize after being cooled to a certain extent,
which results in both a release of latent heat and gravitational energy,
the former being a result of the phase separation of carbon and oxygen
(e.g. Garc\'{\i}a-Berro et al. 2000, 2011; Isern et al. 2000). This release
of latent heat and gravitational energy should induce a cooling time delay,
leading to higher ignition densities (e.g. Lesaffre et al. 2006).

In Fig.\,13, we show the distribution of
WD explosion mass versus WD cooling time before mass accretion.
From this figure it is clear that, for equivalent WD explosion masses, there can
exist a wide range of pre-accretion cooling times.
Interestingly, we find a systematic change in the distribution of median WD explosion masses
for different cooling times (see the red stars in this figure), in
which the peak of the median WD explosion mass is at relatively
\emph{long} cooling times. In Fig.\,14, we present the distribution of
WD explosion mass versus the delay time of SNe Ia that is
corresponding to population age at the time of the SN explosion.  Only
in young populations is the median WD explosion mass significantly in
excess of the Chandrasekhar mass, with a median explosion mass above
1.7\,$\rm M_{\odot}$ for a delay time of 300\,Myr. The explosions from the highest-mass WDs
occur at intermediate ages, with a clear decrease in the frequency of
high-mass explosions at late times (reminiscent of, e.g. Fig. 8 of
Howell et al.\,2009).  Note that the delay times in Fig.\,14 neglect the
potentially long spin-down time for WDs in solid-body rotation.
In Fig. 15, we show the WD cooling time before mass accretion versus SN Ia delay time.
The SNe with long delay times are naturally correlated with long cooling times of WDs.
Both the delay time and WD cooling time are both related to the mass of
the MS star; a low-mass MS star takes a long evolutionary time to fill
its Roche lobe, resulting in a long WD cooling time and a long delay
time (see also Meng, Yang \& Li 2010). However, the differences in the distributions of
median WD masses in Figs\,13$-$14 demonstrate that this is not a
trivial relationship, presumably because \emph{some} WDs with short
cooling times explode after relatively long delay times, i.e. the
formation of the WD was fine-tuned to the time of Roche-lobe filling.

We have shown that, if we allow mass accretion to delay the explosion of the WD, but for otherwise
standard assumptions, most SD SN Ia progenitor systems still produce WDs which have masses at explosion
sufficiently close to the canonical Chandrasekhar limit
for them to avoid explosion whilst remaining in solid-body rotation.  Some systematic diversity may arise from the
pre-explosion cooling, as above. In addition, since a very wide-range of spin-down times seems to be a priori reasonable for
these systems then, in principle, this narrow mass range of systems could produce a diverse range of SNe Ia,
especially if some such WDs are able to crystallize before they explode.  However, since the model predicts
that the majority of SN Ia progenitors fall in this solid-body regime, and since most SNe Ia are inferred to
have masses close to the Chandrasekhar mass (e.g. Mazzali et al.\ 2007), we strongly expect that the ``normal"
range of SNe Ia emerges from within this set of
progenitors. Unfortunately, we cannot accurately predict the variation of explosion masses
for the population of WDs in solid-body rotation. However, we expect that this accounts for at least
some of the observed scatter, and very likely some of the observed systematic diversity.
Note that Dom\'{\i}nguez et al. (2006) has already
explored the explosion of solid-body rotating WDs (up to 1.5\,$\rm M_{\odot}$).

Clearly the extreme ``super-Chandrasekhar" SNe Ia must, if they have SD progenitors, be supported by differential
rotation before explosion.  Assuming that the timescale of the angular momentum redistribution is not unexpectedly long,
post-accretion cooling times seem unlikely to affect the diversity of the explosions produced by this set of
differentially-rotating WDs.  Here, systematic brightness diversity might be related with the wide range of
final WD masses predicted by the model. Note also that the distribution of internal angular-momentum distributions
at explosion (and therefore any dynamo-generated magnetic fields) should be correlated with WD mass.

\section{Summary}\label{7. SUMMARY}

We have performed detailed binary evolution calculations of the WD +
MS channel for the formation of SNe Ia. In an improvement over our
previous works, we have allowed the WD to exceed the Chandrasekhar mass
due to the accretion of angular momentum.  We then combined these results
with our BPS code to provide estimates for the population of SN Ia
birthrates and delay times, along with the properties of the systems at explosion.
The Galactic SN Ia birthrates from this work are $\sim$$1.36\times 10^{-3}\ {\rm yr}^{-1}$
based on our standard model, and the delay times are $\sim$250\,Myr$-$1.4\,Gyr.
The birthrates from Hachisu et al. (2012b) are higher than those of this work
due to a different method adopted.\footnote{Hachisu et al. (2012b) used
Equation (1) of Iben \& Tutukov (1984) to estimate the birthrate of SNe Ia.
However, Wang et al. (2009b) found that the SN Ia birthrate can be overestimated
by this method.}

For our standard model, we find that 77\% SNe Ia have WD explosion masses in the range
of 1.378--1.5$\,\rm M_{\odot}$, which is approximately the range of WD
masses for which solid-body rotation is, in principle, able to delay
the explosion of the WD.  For WDs with initial mass $<$0.9$\,\rm M_{\odot}$,
all of the final WD masses are predicted to be below 1.5$\,\rm M_{\odot}$.
This high fraction of WDs which are predicted to end their
accretion phase with masses $\leq$$1.5\,\rm M_{\odot}$ supports the
suggestion by Justham (2011) that accounting for the spin of the WD
might enable the majority of SD SN Ia progenitors to avoid showing
signs of H contamination (see also Di Stefano, Voss \& Claeys 2011).
In contrast to Hachisu et al.\ (2012b), we are able to accommodate the
population of 91T-like SNe Ia within the set of WDs which explode with
masses $\leq$$1.5\,\rm M_{\odot}$. Some of the diversity of SN
Ia explosion properties might be explained within this range of WD
explosion masses, and the WD spin-down time before SN explosion is also
expected to cause the diversity. Systematic correlations between
population age and SN Ia luminosity might also be partly explained by a
combination of both pre- and post-accretion cooling times.

Some models for the observed extremely luminous SNe Ia invoke the explosion of
WDs with masses far in excess of the Chandrasekhar mass (of
$\sim$$2.0\,\rm M_{\odot}$ or more). Whilst our assumptions about
accretion efficiency may be less reliable in this regime, our
prediction is that 2\% of SNe Ia from the WD + MS channel occur
with masses $\geq$$2.0\,\rm M_{\odot}$, which is broadly comparable with the inferred
rate of these SNe. The proportion from Hachisu et al. (2012b) is
3.8\% that is about two times of ours, but they have a small fraction
with WD explosion masses $\leq$$1.5\,\rm M_{\odot}$ (for the reason see Section 6.1).

For the 23\% of systems which the model predicts to explode with WD
masses $>$1.5$\,\rm M_{\odot}$, we note that the fact that they need to be
supported by differential rotation means that the explosion may well
occur when the WD still has a H-rich companion, and in
systems where the mass transfer is still ongoing.  This assumes that
internal angular momentum redistribution is relatively rapid, which is
broadly expected. The work of Saio \& Nomoto (2004) indicates that
the timescale of angular momentum redistribution could be extremely short
as the angular momentum transport is very fast  (see also Piro 2008).\footnote{The upper limits for
the angular momentum redistribution time might be $\sim$$10^8$\,yr owing to
angular momentum transport by the Eddington-Sweet meridional circulation
(see Yoon \& Langer 2004;  Hachisu et al. 2012a). }
We therefore suggest that these systems may help to
explain the existence of SNe Ia with circumstellar material. In
extreme cases, they might also contribute to the population of type
IIn SNe.

If any SD scenario is correct, then the mass-donating star in the SD model
would survive after the SN explosion, in which case our models may be
useful for constraining the searches for remnant donors. If the spin-up/spin-down
model is the correct scenario for SD SNe Ia, then some pre-explosion SN progenitor
systems should be observable which contain rapidly-rotating WDs in
excess of the Chandrasekhar mass. These could be systems which are continuing
to accrete, with properties similar to those, as noted by Hachisu et al. (2012b),
of U Scorpii, RS Ophiuchi and V445 Puppis. Alternatively, they could be post-accretion systems
in which the WD is spinning down. We hope that observers will find
such post-accretion systems, e.g. to find some binaries with massive WDs that are
close to or above the Chandrasekhar mass, which would provide outstanding
constraints on our understanding of SN Ia progenitors.

\section*{Acknowledgments}
We acknowledge the anonymous referee for valuable comments that helped us
to improve the paper. We thank Paolo Mazzali for discussions about the Zorro
diagram and inferring WD pre-explosion masses. We also thank Philipp Podsiadlowski,
Xiangdong Li, Sung-Chul Yoon, Xiangcun Meng and Xuefei Chen for their helpful discussions.
This work is supported by the National Basic Research Program of
China (No. 2014CB845700), the National Natural Science Foundation of
China (Nos. 11322327, 11103072, 11033008, 11390374, 11250110055 and 11350110324),
and the Natural Science Foundation of Yunnan Province (Nos. 2013FB083 and 2013HB097).

\label{lastpage}
\end{document}